\documentclass[preprint,superscriptaddress,showpacs,tightenlines]{revtex4}
\usepackage{amssymb}
\usepackage{amsmath}
\usepackage{graphicx,bm}

\input{tcilatex}

\begin{document}

\title{Magneto-quantum oscillations of the conductance of a tunnel
point-contact in the presence of a single defect.}
\author{Ye.S. Avotina}
\affiliation{B.I. Verkin Institute for Low Temperature Physics and Engineering, National
Academy of Sciences of Ukraine, 47, Lenin Ave., 61103, Kharkov,Ukraine.}
\affiliation{Kamerlingh Onnes Laboratorium, Universiteit Leiden, Postbus 9504, 2300
Leiden, The Netherlands.}
\author{Yu.A. Kolesnichenko}
\affiliation{B.I. Verkin Institute for Low Temperature Physics and Engineering, National
Academy of Sciences of Ukraine, 47, Lenin Ave., 61103, Kharkov,Ukraine.}
\affiliation{Kamerlingh Onnes Laboratorium, Universiteit Leiden, Postbus 9504, 2300
Leiden, The Netherlands.}
\author{A.F. Otte}
\affiliation{Kamerlingh Onnes Laboratorium, Universiteit Leiden, Postbus 9504, 2300
Leiden, The Netherlands.}
\author{J.M. van Ruitenbeek}
\affiliation{Kamerlingh Onnes Laboratorium, Universiteit Leiden, Postbus 9504, 2300
Leiden, The Netherlands.}

\begin{abstract}
The influence of a strong magnetic field $H$ to the conductance of
a tunnel point contact in the presence of a single defect has been
considered. We demonstrate that the conductance exhibits specific
magneto-quantum oscillations, the amplitude and period of which
depend on the distance between the contact and the defect. We show
that a non-monotonic dependence of the point-contact conductance
results from a superposition of two types of oscillations: A short
period oscillation arising from the electrons being focused by the
field $H$ and a long period oscillation originated from the
magnetic flux passing through the closed trajectories of electrons
moving from the contact to the defect and returning back to the
contact.
\end{abstract}

\maketitle

\section{Introduction}

The presence of a single defect in the vicinity of a point contact
manifests itself in an oscillatory dependence of the conductance
$G$ on the applied voltage $V$ and the distance between the
contact and the defect. Conductance oscillations originate from
quantum interference between electrons that pass directly through
the contact and electrons that are backscattered by the defect and
again forward scattered by the contact. The reason of the
oscillations of $G(V)$ is a dependence of the phase shift between
two waves on the electron energy, which depends on the bias $eV$.
This effect has been observed experimentally
\cite{Ludoph1,Untiedt,Ludoph,Kempen} and investigated
theoretically \cite{Avotina1,Namir,Avotina2,Avotina3,Avotina4}. In
an earlier paper \cite{Avotina1} we demonstrated that this $G(V)$
dependence can actually be used to determine the exact location of
a defect underneath a metal surface by means of scanning
tunnelling microscopy (STM). A more elaborate version of this
method \cite{Avotina4} that takes the Fermi surface anisotropy
into account corresponds quite well with experimental observations
\cite{Wend}. Here we consider another way to change the phase
shift between the interfering waves: By applying an external
magnetic field $\mathbf{H}$ we expect to observe oscillations of
the conductance as a function of the field $\mathbf{H.}$

It is well known that a high magnetic field $\mathbf{H}$
fundamentally changes the kinetic and thermodynamic
characteristics of a metal \cite{LAK,Abrikosov}. When speaking of
a high magnetic field one usually assumes two conditions to  be
fulfilled. The first one is that the radius of the electron
trajectory, $r_{H},$ is much smaller than the mean free path of
electrons, $l$. This condition implies that electrons move along
spiral trajectories between two scattering events, such as by
defects or phonons. This change in character of the electron
motion results, for example, in the phenomenon of
magnetoresistance \cite{LAK,Abrikosov}. The second condition
requires that the distance between the magnetic quantum levels,
the Landau levels, $\hbar \Omega $ ($\Omega $ is the frequency of
the electron motion in the magnetic field $H$) is larger than the
temperature $k_{\rm B}T.$ Under this condition oscillatory quantum
effects, such as the de Haas-van Alphen and Shubnikov-de Haas
oscillations, can be observed \cite{LAK,Abrikosov}. At which
actual value the field $H$  can be identified as a high depends on
the purity of the metal, its electron characteristics and the
temperature of the experiment. Typically, the high field condition
requires fields values above 10T for metals at low temperatures,
$T\simeq 1$K,  while for a pure bismuth monocrystal (a semimetal)
a field of $H\simeq 0.1$T is sufficient to satisfy the two
conditions mentioned.

A high magnetic field $\mathbf{H}$ influences the current
spreading of the electrons passing through the contact. If the
vector $\mathbf{H}$ is parallel to the contact axis, the electron
motion becomes quasi-one-dimensional. Electrons then move inside a
`tube' with a diameter defined by the contact radius, $a$, and the
radius $r_{H}$. The three-dimensional spreading of the current is
restored by elastic and inelastic scattering processes. As shown
in \cite{Bogachek}, for $r_{H}\ll a$ and $r_{H}\ll l$, the contact
resistance increases linearly with the magnetic field, in contrast
to bulk samples for which the resistance increases as $H^{2}$. The
Shubnikov-de Haas oscillations in the resistance of `large'
contacts (defined by $a\gg \lambda _{\mathrm{F}}$, with $\lambda
_{\mathrm{F}}$ the electron Fermi wave length) were considered
theoretically in Refs.~\cite{Bogachek1,Bogachek2}. Experimentally,
a point-contact magnetoresistance linear in $H$ as well as
Shubnikov-de Haas oscillations were observed for bismuth
\cite{Andr}.

In this paper we consider the influence of a high magnetic field
on the linear conductance (Ohm's law approximation, $V\rightarrow
0$) of a tunnel point contact in the presence of a single defect,
with the magnetic field directed along the contact axis. We
demonstrate that the conductance exhibits magneto-quantum
oscillations, the amplitude and period of which depend on the
distance between the contact and the defect. We show that the
non-monotonic dependence of the conductance $G(H)$ results from
the superposition of two types of oscillations: (a) A short period
oscillation arising from electrons being focused by the field $H$
and (b) a long period oscillation of Aharonov-Bohm-type
originating from the magnetic flux passing through the area
enclosed by the electron trajectories from contact to defect and
vice versa.

In Sec.~II we will discuss the model of a tunnel point-contact and find the
electron wave function in the limit of a high potential barrier at the
contact. The interaction of the electrons with a single impurity placed
nearby the contact is taken into account by perturbation theory, with the
electron-impurity interaction as the small parameter. A general analytical
expression for the dependence of conductance, $G(H)$, on the magnetic field $%
H$ is obtained in Sec.~III. It describes $G(H)$ in terms of the distance
between contact and defect and the value of the magnetic field. The physical
interpretation of of the general expression for the conductance can be
obtained from the semiclassical asymptotics given in this Section. In
Sec.~IV we conclude by discussing our results and the feasibility of finding
the predicted effects experimentally.

\section{Model and electron wave function}

\begin{figure}[tbp]
\includegraphics[width=10cm,angle=0]{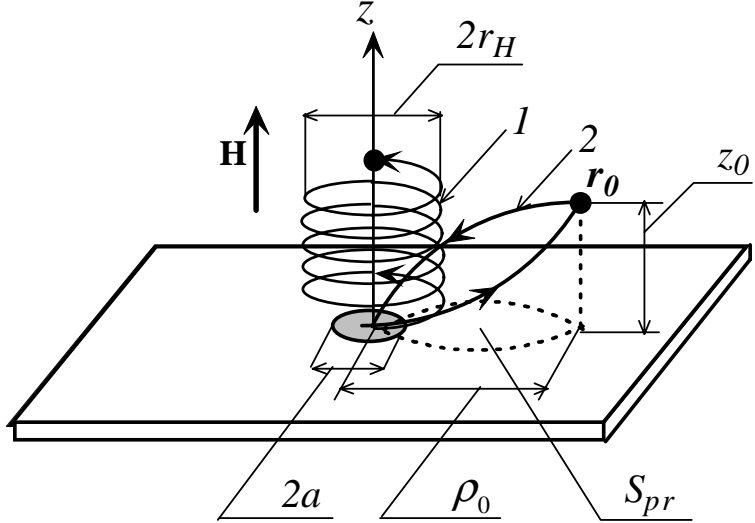}
\caption{Model of a tunnel point-contact. The upper and lower
metal half-spaces are separated by an inhomogeneous barrier
(Eq.~(\protect\ref{U})) that allows electron tunneling mainly in a
small region with a characteristic radius $a$, which defines the
tunneling point contact. A
single defect is placed inside the upper metal at the position $\mathbf{r}%
_{0}$. Electron trajectories in the magnetic field are shown schematically.}
\label{Fig-model}
\end{figure}

Let us consider a point-contact centered at the point $\mathbf{r}=0$, as
illustrated in Fig.~\ref{Fig-model}.
We use cylindrical coordinates $\mathbf{r}=\left( \rho ,\varphi ,z\right) $
with the $z$-axis directed along the axis of the contact. The potential
barrier in the plane $z=0$ is taken to be defined by a $\delta $-function of
the form,
\begin{equation}
U(\rho ,\varphi ,z)=Uf(\rho )\delta (z).  \label{U}
\end{equation}%
In order to allow for the current to flow only through a small region near
the point $\mathbf{r}=0$ we choose the model function
\begin{equation}
f(\rho )=e^{\rho ^{2}/a^{2}},  \label{f-rho}
\end{equation}%
where the small $a$ specifies the characteristic radius of the contact. A
point-like defect is placed at the point $\mathbf{r}=\mathbf{r}_{0}$ in
vicinity of the interface in the half-space $z>0$, see Fig.~\ref{Fig-model}.
The scattering of electrons with the defect is described by a potential $%
D(\left\vert \mathbf{r}-\mathbf{r}_{0}\right\vert )$, which is localized
near the point $\mathbf{r}=\mathbf{r}_{0}$ in a small region with a
characteristic radius, which is of the order of the Fermi wave length $%
\lambda _{\mathrm{F}}.$ The screened Coulomb potential is an
example of such kind of dependence of $D( r) $ \cite{Kittel}. It
is widely used to describe charge point defects (impurities) in
metals.

We assume that the transmission probability of electrons through
the barrier, Eq.~(\ref{U}), is small such that the applied voltage
drops entirely over the
barrier. We can then take the electric potential as a step function $%
V(z)=V\,\Theta (-z)$. The magnetic field is directed along the contact axis,
$\mathbf{H}=(0,0,H)$. In cylindrical coordinates the vector-potential $%
\mathbf{A}$ has components $A_{\varphi }=H\rho /2,$ $A_{z}=A_{\rho }=0.$

The Schr\"{o}dinger equation for the wave function $\psi (\rho \mathbf{,}%
\varphi ,z)$ is given by%
\begin{gather}
-\frac{\hbar ^{2}}{2m^{\ast }}\left[ \frac{1}{\rho }\frac{\partial }{%
\partial \rho }\left( \rho \frac{\partial \psi }{\partial \rho }\right) +%
\frac{\partial ^{2}\psi }{\partial z^{2}}+\frac{1}{\rho ^{2}}\frac{\partial
^{2}\psi }{\partial \varphi ^{2}}\right] -\frac{i\hbar \Omega }{2}\frac{%
\partial \psi }{\partial \varphi }+  \label{Schrod} \\
+\left( \frac{m^{\ast }\Omega ^{2}}{8}\rho ^{2}+U(\rho \mathbf{,}z)+D(\rho
\mathbf{,}\varphi ,z)\right) \psi =\left( \varepsilon +\sigma \mu _{\mathrm{B%
}}H\right) \psi ,  \notag
\end{gather}%
where $\Omega =eH/m^{\ast }c$; $\varepsilon $ and $m^{\ast }$ are
the electron energy and effective mass, respectively, and $e$ is
the absolute value of the electron charge, $\sigma =\pm 1$
corresponds to different spin directions, $\mu
_{\mathrm{B}}=e\hbar /2m_{0}c$ is the Bohr magneton, where $m_{0}$
is the free electron mass. Hereinafter assuming that $\mu _{\mathrm{B}%
}H/\varepsilon _{\mathrm{F}}\simeq \lambdabar _{\mathrm{F}}/r_{H}\ll 1$ we
will neglect by the term $\sigma \mu _{\mathrm{B}}H$ in Eq.(\ref{Schrod}).

In order to solve Eq.~(\ref{Schrod}) in the limit of a high potential
barrier we use the method that was developed in Refs.~\cite{KMO,Avotina1}.
To first order approximation in the small parameter $\hbar p_{z}/m^{\ast
}U\ll 1$, which leads to a small electron tunnelling probability $T\approx
\left( \hbar p_{z}/m^{\ast }U\right) ^{2}\ll 1$, the wave function $\psi $
can be written in the form,
\begin{eqnarray}
\psi (\rho ,\varphi ,z) &=&\psi _{0}(\rho ,\varphi ,z)+\varphi ^{(-)}(\rho
,\varphi ,z)\qquad \,\,(z<0),  \label{psi<0} \\
\psi (\rho ,\varphi ,z) &=&\varphi ^{(+)}(\rho ,\varphi ,z)\qquad \qquad
\qquad \qquad (z>0),  \label{psi>0}
\end{eqnarray}%
where $\psi _{0}$ does not depend on $U$, but $\varphi ^{(\pm )}\sim 1/U.$
In Eq.~(\ref{psi<0}) $\psi _{0}$ is the wave function in the absence of
tunnelling, for $U\rightarrow \infty $. It satisfies the boundary condition $%
\psi _{0}(\rho ,\varphi ,0)=0$ at the interface. Using the well known
solution of the Schr\"{o}dinger equation for an electron in a magnetic field
\cite{LandauQM} the energy spectrum and wave function $\psi _{0}$ are given
by,
\begin{equation}
\varepsilon =\varepsilon _{mn}+\frac{p_{z}^{2}}{2m^{\ast }},\qquad
\varepsilon _{mn}=\hbar \Omega \left( n+\frac{m+\left\vert m\right\vert +1}{2%
}\right) ,
\end{equation}%
\begin{equation}
\psi _{0}(\rho ,\varphi ,z)=e^{im\varphi }\left( e^{\frac{i}{\hbar }%
p_{z}z}-e^{-\frac{i}{\hbar }p_{z}z}\right) R_{nm}(\rho ),
\end{equation}%
where
\begin{equation}
R_{nm}(\rho )=\left[ \frac{\left( n\right) !}{\left( \left\vert m\right\vert
+n\right) !}\right] ^{1/2}\exp \left( -\frac{\xi }{2}\right) \xi
^{\left\vert m\right\vert /2}L_{n}^{\left\vert m\right\vert }(\xi ).
\label{Rnm}
\end{equation}%
Here, $\xi =\rho ^{2}/2a_{H}^{2}$, and $L_{n}^{\left\vert
m\right\vert }(\xi ) $ are the generalized Laguerre polynomials,
$a_{H}=\sqrt{\hbar /m^{\ast
}\Omega }$ is the quantum magnetic length, $n=0,1,2...,$ $m=0,\pm 1,\pm 2...$%
, and $p_{z}$ is the electron momentum along the vector $\mathbf{H}$. The
functions (\ref{Rnm}) are orthogonal. We use a normalization of the wave
function (\ref{Rnm}) such that $R_{n0}(0)=1.$

The function $\varphi ^{(-)}(\rho ,\varphi ,z)$ in Eq.~(\ref{psi<0})
describes the correction to the reflected wave as a result of a finite
tunnelling probability and $\varphi ^{(+)}(\rho ,\varphi ,z)$, Eq.~(\ref%
{psi>0}), is the wave function for the electrons that are transmitted
through the contact. The wave functions (\ref{psi<0}) and (\ref{psi>0})
should be matched at the interface $z=0.$ For large $U$ the resulting
boundary conditions for the functions $\varphi ^{(-)}$ and $\varphi ^{(+)}$
become \cite{KMO},
\begin{gather}
\varphi ^{(-)}(\rho ,\varphi ,0)=\varphi ^{(+)}(\rho ,\varphi ,0);
\label{bound1} \\
ip_{z}=\frac{m^{\ast }U}{\hbar }f(\rho )\varphi ^{(+)}(\rho ,\varphi ,0).
\label{gran_cond}
\end{gather}%
In order to proceed with further calculations we assume that the
electron-impurity interaction is small and use perturbation theory \cite%
{Avotina1}. In the zeroth approximation in the defect scattering potential
the function $\varphi _{0}^{(+)}$ can be found by means of the expansion of
the function $\varphi _{0}^{(+)}(\rho ,\varphi ,0)$ over the full set of
orthogonal functions $R_{nm}(\rho )$, Eq.~(\ref{Rnm}), and $\varphi
_{0}^{(+)}(\rho ,\varphi ,z)$ is given by
\begin{equation}
\varphi _{0}^{(+)}(\rho ,\varphi ,z)=-\frac{i\hbar p_{z}}{m^{\ast }U}\frac{1%
}{2\pi a_{H}^{2}}\sum\limits_{n^{\prime }=0}^{\infty }F_{nn^{\prime
},m}e^{im\varphi }R_{n^{\prime }m}(\rho )\exp \left( \frac{i}{\hbar }%
p_{z,n^{\prime }m}z\right) ,\text{ \ }
\end{equation}%
for $z\neq 0$. Here,%
\begin{equation}
p_{z,nm}=\sqrt{2m^{\ast }\left( \varepsilon -\varepsilon _{nm}\right) },
\label{pz}
\end{equation}%
and
\begin{equation}
F_{nn^{\prime },m}=\int\limits_{0}^{a}d\rho \;\rho f(\rho )R_{nm}(\rho
)R_{n^{\prime }m}^{\ast }(\rho ).  \label{F}
\end{equation}%
For the model function $f(\rho )$ of Eq.~(\ref{f-rho}) the integral (\ref{F}%
) can be evaluated and the function $F_{nn^{\prime },m}$ takes the form
\begin{eqnarray}
F_{nn^{\prime },m} &=&\left[ \frac{\left( \left\vert m\right\vert +n\right)
!\left( \left\vert m\right\vert +n^{\prime }\right) !}{\left( n\right)
!\left( n^{\prime }\right) !}\right] ^{1/2}\frac{\pi a^{2}}{\left(
\left\vert m\right\vert \right) !}\left( \frac{a^{2}}{2a_{H}^{2}}\right)
^{\left\vert m\right\vert }\left( 1-\frac{a^{2}}{2a_{H}^{2}}\right)
^{\left\vert m\right\vert +1+n+n^{\prime }}\times  \label{F1} \\
&&_{2}F_{1}\left( \left\vert m\right\vert +1+n^{\prime },\left\vert
m\right\vert +1+n,\left\vert m\right\vert +1,\frac{a^{4}}{4a_{H}^{4}}\right)
,  \notag
\end{eqnarray}%
where $_{2}F_{1}(a,b,c,\xi )$ is a hypergeometric function. By using the
procedure developed in the Ref. \cite{Avotina1} we find the wave function $%
\varphi ^{(+)}(\rho ,\varphi ,z)$ at $z>z_{0}$ accurate to $g$%
\begin{eqnarray}
\varphi ^{(+)}(\rho ,\varphi ,z) &=&\varphi _{0}^{(+)}(\rho ,\varphi ,z)+%
\frac{im^{\ast }g}{2\pi \hbar }\frac{1}{2\pi a_{H}^{2}}\varphi
_{0}^{(+)}(\rho _{0},\varphi _{0},z_{0})\times  \label{psi0+g} \\
&&\sum\limits_{n^{\prime }=0}^{\infty }\sum\limits_{m^{\prime }=-\infty
}^{\infty }\frac{e^{im^{\prime }\left( \varphi -\varphi _{0}\right) }}{%
p_{z}^{\prime }}R_{n^{\prime }m^{\prime }}(\rho )R_{n^{\prime }m^{\prime
}}^{\ast }(\rho _{0})\left( e^{\frac{i}{\hbar }p_{z}^{\prime }\left(
z-z_{0}\right) }-e^{\frac{i}{\hbar }p_{z}^{\prime }\left( z+z_{0}\right)
}\right) ,  \notag
\end{eqnarray}%
where%
\begin{equation}
g=\int d\mathbf{r}^{\prime }D(\left\vert \mathbf{r}^{\prime }-\mathbf{r}%
_{0}\right\vert )  \label{g}
\end{equation}%
is the interaction constant for the scattering of the electron with the
impurity. We proceed in Sec.~III to calculate the total current through the
contact and the point-contact conductance, using the wave function (\ref%
{psi0+g}).

\section{Total current and point-contact conductance}

The electrical current $I(H)$ can be evaluated from the electron
wave functions of the system, $\psi $ \cite{ISh}. We shall also
assume that the applied bias $eV$ is much smaller than the Fermi
energy, $\varepsilon _{\mathrm{F}}$, and calculate the conductance
in linear approximation in $V$. In this approximation we find
\begin{equation}
I(H)=-\frac{2\left\vert e\right\vert ^{3}HV}{\left( 2\pi \hbar \right) ^{2}c}%
\sum\limits_{n=0}^{\infty }\sum\limits_{m=-\infty }^{\infty }\int
dp_{z}I_{nm,p_{z}}\Theta (p_{z})\frac{\partial n_{\mathrm{F}}(\varepsilon )}{%
\partial \varepsilon }.  \label{current}
\end{equation}%
Here%
\begin{equation}
I_{nm,p_{z}}=\frac{\hbar }{m^{\ast }}\int\limits_{0}^{2\pi }d\varphi
\int\limits_{0}^{\infty }\rho \,d\rho \ \mathrm{\func{Re}}\left( \left(
\varphi ^{(+)}\right) ^{\ast }\frac{\partial \varphi ^{(+)}}{\partial z}%
\right)   \label{den_flow}
\end{equation}%
is the probability current density in the $z$ direction, integrated over
plane $z=$ constant; $n_{\mathrm{F}}(\varepsilon )$ is the Fermi
distribution function. For a small contact, $a\ll a_{H}$, Eq.~(\ref{current}%
) can be simplified. The largest term in the parameter $a^{2}/2a_{H}^{2}\ll 1
$ in Eq.~(\ref{current}) corresponds to $m=0$, for which the Eq.~(\ref{F1})
takes the form $F_{nn^{\prime },0}\thickapprox \pi a^{2}$.

After space integration over a plane at $z>z_{0},$ where the wave function (%
\ref{psi0+g}) can be used, we obtain the current density (\ref{den_flow}).
At low temperatures, $T\rightarrow 0$, the integral over $p_{z}$ in Eq.~(\ref%
{current}) can be easily calculated. The point-contact conductance, $G$, is
the first derivative of the total current $I$ over the voltage $V:$
\begin{eqnarray}
G(H) &=&G_{c}\left( 1+\frac{gm^{\ast }}{4\pi ^{3}N(\varepsilon _{\mathrm{F}%
})\hbar ^{2}a_{H}^{4}}\func{Im}\left( \sum\limits_{n^{\prime }=0}^{n_{\max
}(\varepsilon _{\mathrm{F}})}\chi (\varepsilon _{\mathrm{F}},n^{\prime },%
\mathbf{r}_{0})\right) \right. \times  \label{G} \\
&&\left. \func{Re}\sum\limits_{n^{\prime \prime }=0}^{\infty }\chi
(\varepsilon _{\mathrm{F}},n^{\prime \prime },\mathbf{r}_{0})\right) .
\notag
\end{eqnarray}%
Here
\begin{equation}
\chi (\varepsilon ,n,\mathbf{r}_{0})=R_{n0}(\rho _{0})\exp \left( \frac{i}{%
\hbar }p_{z,n0}z_{0}\right) ,  \label{khi}
\end{equation}%
$N(\varepsilon )$ is the number of electron states per unit volume,%
\begin{equation}
N(\varepsilon ,H)=\frac{4\left\vert e\right\vert H}{\left( 2\pi \hbar
\right) ^{2}c}\sum\limits_{n=0}^{n_{\max }(\varepsilon )}\sqrt{2m\left(
\varepsilon -\varepsilon _{n0}\right) },
\end{equation}%
and $n_{\max }(\varepsilon )=\left[ \frac{\varepsilon }{\hbar \Omega }\right]
$ is the maximum value of the quantum number $n$ for which $\varepsilon
_{n0}<\varepsilon ,$ and $\left[ x\right] $ is the integer part of the number $%
x. $ The constant $G_{c}$ is the conductance in absence of a
defect,
\begin{equation}
G_{c}( H) =\pi ^{3}e^{2}\hbar ^{3}\left( \frac{a^{2}N(\varepsilon
_{\mathrm{F}})}{2m^{\ast }U}\right) ^{2}. \label{G0}
\end{equation}

\begin{figure}[tbp]
\includegraphics[width=10cm,angle=0]{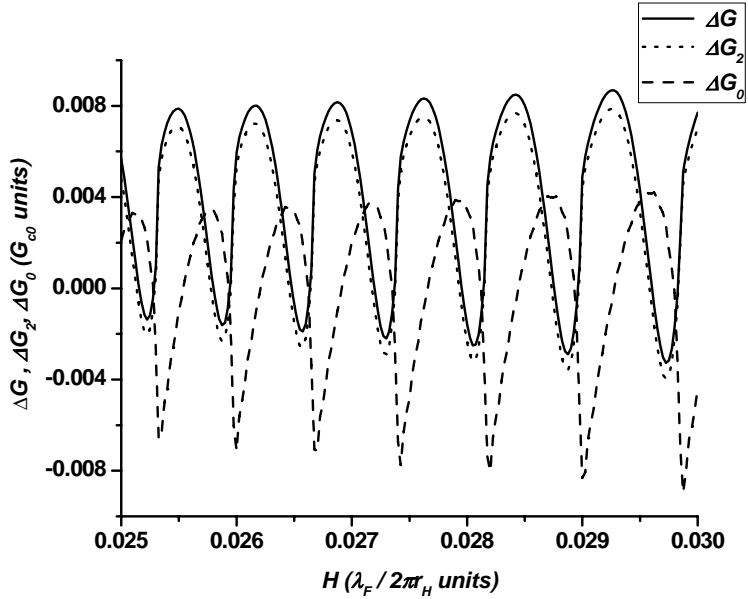}
\caption{Oscillatory part of the conductance for a defect placed
on the contact axis, $\protect\rho _{0}=0,$ $z_{0}=30\lambdabar
_{\mathrm{F}}$. The full curve is a plot for
Eq.~(\protect\ref{G}), while the dotted curve shows the component
$\Delta G_{2}$ for the semiclassical approximation,
Eq.~(\protect\ref{dG2}), and the dashed curve shows the component
$\Delta G_{0}$, Eq. (\protect\ref{dG0}). The constant of
electron-defect interaction is
taken as $\widetilde{g}=0.5.$ The field scale is given in units $\lambdabar _{%
\mathrm{F}}/r_{H}=(e\hbar /p_{\mathrm{F}}^{2}c)H$. }
\label{Fig-G_osc 0}
\end{figure}

\begin{figure}[tbp]
\includegraphics[width=10cm,angle=0]{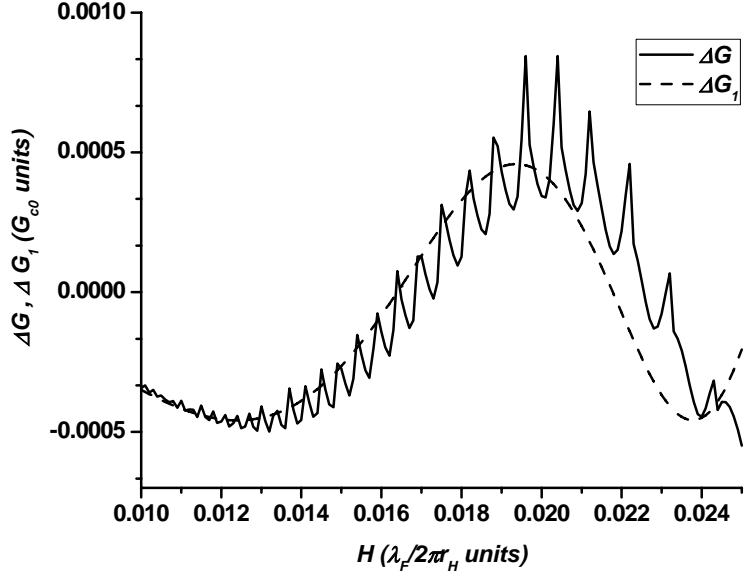}
\caption{Oscillatory part of the conductance of a tunnelling point contact
with a single defect placed at $\protect\rho _{0}=50\lambdabar _{\mathrm{F}},
$ $z_{0}=30\lambdabar _{\mathrm{F}}.$ The full curve is a plot for Eq.~(%
\protect\ref{G}), while the dashed curve shows the component $\Delta G_{1}$
for the semiclassical approximation, Eq.~(\protect\ref{dG1}). The field
scale is given in units $\lambdabar _{\mathrm{F}}/r_{H}=(e\hbar /p_{\mathrm{F%
}}^{2}c)H$; $\widetilde{g}=0.5$. }
\label{Fig-G_osc}
\end{figure}

The second term in brackets in Eq.~(\ref{G}) describes the
oscillatory part of the conductance, $\Delta G(H)=G(H) -G_{c}(H) $
that results from the scattering by the defect. This term is
plotted in Fig.~\ref{Fig-G_osc 0} for a defect placed on the
contact axis (solid curve). We find an oscillatory dependence
which is dominated by a single period, although the shape is not
simply harmonic. However, this dependence becomes quite
complicated and contains oscillations having different periods
when the defect is not sitting on the contact axis, as illustrated
by the example plotted in Fig.~\ref{Fig-G_osc} (solid curve) for a
defect placed at
$(\rho ,z)=(50,30)$ (in units $\lambdabar _{\mathrm{F}}=\hbar /p_{\mathrm{F}%
},$ with $p_{\mathrm{F}}=\sqrt{2m^{\ast }\varepsilon
_{\mathrm{F}}}$ the Fermi momentum) . The physical origin of the
oscillations can be extracted from the semiclassical asymptotics
of Eq~.(\ref{G}).

For magnetic fields that are not too high one typically has a large number
of Landau levels, $n_{\max }(\varepsilon _{\mathrm{F}})\approx \varepsilon _{%
\mathrm{F}}/\hbar \Omega =\left( a_{H}/\sqrt{2}\lambdabar _{\mathrm{F}%
}\right) ^{2}\gg 1$, in which case the semiclassical approximation
can be used. Some details of the calculations are presented in the
Appendix. The asymptotic form of the expression for the conductance Eq.~(\ref{G}%
) can be written as a sum of four terms
\begin{equation}
G(H)=G_{c0}+\Delta G_{0}+\Delta G_{1}+\Delta G_{2}.  \label{sum}
\end{equation}%
In leading approximation in the small parameter $\hbar \Omega /\varepsilon _{%
\mathrm{F}}$ the conductance (\ref{G0}) does not depend on the magnetic
field
\begin{equation}
G_{c0}=\frac{4e^{2}}{9\pi \hbar }T(p_{\mathrm{F}})\left( \frac{p_{\mathrm{F}%
}a}{\hbar }\right) ^{4},  \label{G0_asym}
\end{equation}%
where $T(p_{\mathrm{F}})=\left( \hbar p_{\mathrm{F}}/m^{\ast }U\right)
^{2}\ll 1$ is the transmission coefficient of the tunnel junction. There is
an oscillatory contribution $\Delta G_{0}$ to the conductance that
originates from the step-wise dependence of the number of states $%
N(\varepsilon _{\mathrm{F}})$ on the magnetic field, and the conductance
undergoes oscillations having the periodicity of the de Haas-van Alphen
effect,
\begin{equation}
\Delta G_{0}=\frac{9}{2}G_{c0}\left( \frac{\lambdabar _{\mathrm{F}}}{a_{H}}%
\right) ^{3}\sum\limits_{k=1}^{\infty }\frac{\left( -1\right) ^{k}}{k^{3/2}}%
\sin \left( \pi k\frac{a_{H}^{2}}{\lambdabar _{\mathrm{F}}^{2}}-\frac{\pi }{4%
}\right) .  \label{dG0}
\end{equation}%
The other two terms in Eq.~(\ref{sum}), $\Delta G_{1}$ and $\Delta G_{2}$,
result from the electron scattering on the defect.

Using the results presented in the Appendix, Eq.~(\ref{S_asym}),
we find for the first oscillation,
\begin{equation}
\Delta G_{1}(H,\mathbf{r}_{0})=-G_{c0}\widetilde{g}\frac{z_{0}^{2}\lambdabar
_{\mathrm{F}}^{2}}{r_{0}^{4}}\sin \left( \frac{2p_{\mathrm{F}}r_{0}}{\hbar }%
-2\pi \frac{\Phi }{\Phi _{0}}\right) .  \label{dG1}
\end{equation}%
where $\widetilde{g}=3gm^{\ast }p_{\mathrm{F}}/4\pi \hbar ^{3}$ is
a dimensionless constant representing the defect scattering
strength, and $\Phi _{0}=2\pi \hbar c/e$ is the flux quantum. The
flux,
\begin{equation}
\Phi =HS_{\mathrm{pr}},  \label{FI}
\end{equation}%
is given by the field lines penetrating the area of the projection
$S_{\mathrm{pr}}=2S_{\mathrm{seg}}$ on the plane $z=0$ of the
trajectory of the electron moving from the contact to the defect
and back (see, Fig.~\ref{Fig-model}),
\begin{equation}
S_{\mathrm{seg}}=r^{2}\left( \theta _{\mathrm{st}}-\sin 2\theta _{\mathrm{st}%
}\right) .
\end{equation}%
$S_{\mathrm{seg}}$ is the area of the segment formed by the chord of length $%
\rho _{0}$ and the arc of radius $r=r_{H}\sin \theta _{\mathrm{st}}$, with $%
\theta _{\mathrm{st}}$ is the angle between the vector $\mathbf{r}_{0}$ and $%
z$-axis, $\sin \theta _{\mathrm{st}}=\rho _{0}/r_{0},$ $r_{H}=cp_{\mathrm{F}%
}/eH.$ The oscillation $\Delta G_{1}$ disappears when the defect sits on the
contact axis, $\rho _{0}=0.$ Note that for $H\rightarrow 0$ Eq.~(\ref{dG1})
reduces to the expression obtained before \cite{Avotina1} for the
point-contact conductance in the presence of a defect.

An analytic
expression for the last term $\Delta G_{2}(H,\mathbf{r}_{0})$ in Eq.~(\ref%
{sum}) can be written by use of Eq.~(\ref{S2_asym}) as
\begin{gather}
\Delta G_{2}(H,\rho _{0}=0,z_{0})=G_{c0}\widetilde{g}\left( \frac{\lambdabar
_{\mathrm{F}}}{a_{H}}\right) ^{3}\times  \label{dG2} \\
\left\{ \sum\limits_{k=\left[ z_{0}/2\pi r_{H}\right] }^{\infty }\left(
-1\right) ^{k}\frac{1}{k^{3/2}}\cos \left( \frac{p_{\mathrm{F}}r_{0}}{\hbar }%
+\pi k\frac{a_{H}^{2}}{\lambdabar _{\mathrm{F}}^{2}}+\frac{z_{0}^{2}}{4\pi
ka_{H}^{2}}\right) \right. +  \notag \\
\frac{z_{0}^{2}\sqrt{2}}{a_{H}^{2}}\left( \frac{\lambdabar _{\mathrm{F}}}{%
a_{H}}\right) \sum\limits_{k,k^{\prime }=\left[ z_{0}/2\pi r_{H}\right]
}^{\infty }\left( -1\right) ^{k+k^{\prime }}\frac{1}{\left( kk^{\prime
}\right) ^{3/2}}\times  \notag \\
\left. \sin \left( \pi k\frac{a_{H}^{2}}{\lambdabar _{\mathrm{F}}^{2}}+\frac{%
z_{0}^{2}}{4\pi ka_{H}^{2}}\right) \cos \left( \pi k^{\prime }\frac{a_{H}^{2}%
}{\lambdabar _{\mathrm{F}}^{2}}+\frac{z_{0}^{2}}{4\pi k^{\prime }a_{H}^{2}}%
\right) \right\} .  \notag
\end{gather}%
As a consequence of the decreasing amplitudes of the summands with $k$ and $%
k^{\prime }$ the main contribution to the conductance oscillations
results from the first term in the braces, with $k=\left[
z_{0}/2\pi r_{H}\right]$. Comparing the dependence $G(H)$ that is
obtained from Eq.(\ref{G}) with the asymptotic expressions
Eq.(\ref{dG2}) in Fig.~\ref{Fig-G_osc 0}, and Eq.(\ref{dG1}) in
Fig.~\ref{Fig-G_osc}, we observe the good agreement between the
exact solution and results obtained in the framework of
semiclassical approximation. This agreement allows us to explain
the nature of the complicated oscillations of the conductance
$G(H).$

\section{Discussion}

The de Haas-van Alphen effect and the Shubnikov-de Haas effect are
quite different manifestations of the Landau quantization of the
electron energy spectrum in a magnetic field. The de Haas-van
Alphen effect is a thermodynamic property that results from
singularities in the electron density of states while the
Shubnikov-de Haas effect is a manifestation of the Landau
quantization due to corrections in the electron scattering
\cite{LAK,Abrikosov}. It is known that a calculation of the
metallic conductivity in a strong magnetic field in the
approximation of a constant mean free scattering time gives an
incorrect answer for the amplitude of the oscillations
\cite{Lifshits}. The correct amplitude can be obtained when the
quantization is taken into account in the collision term of the
quantum kinetic equation \cite{Kosevich}.

We have considered the limiting case when there is only one
scatterer and found specific magneto-quantum oscillations, the
amplitude of which depends on the position of the defect. In our
system a few quantum effects manifest themselves at the same time:
1) the Landau quantization, 2) the quantum interference between
the wave that is directly transmitted through the contact and the
partial wave that is scattered by the contact and the defect, 3)
the effect of the quantization of the magnetic flux. As a
consequence the conductance $G(H)$, Eq.(\ref{G}), is a complicated
non-monotonous function of the magnetic field, see
Figs.~\ref{Fig-G_osc 0} and \ref{Fig-G_osc}.

First of all, Landau quantization results in the oscillations
$\Delta G_{0}(H)$ of Eq.(\ref{dG0}), having the usual period of
the Shubnikov-de Haas (or de Haas-van Alphen) oscillations. From
the point of view of the first paragraph of this section, the
oscillatory part of the conductance (\ref{dG0}) is not a
manifestation of the Shubnikov-de Haas effect but it is due to the
oscillations in the number of states that modify the conductivity
of the tunnel junction.

At $H=0$ the quantum interference between partial electron waves (the
directly transmitted wave and the wave scattered by the defect and reflected
back to the contact) leads to an oscillatory dependence of the conductance
as a function of the position of the defect \cite{Avotina1} and the period
of this oscillation can be found from the phase shift $\Delta \phi =2p_{%
\mathrm{F}}r_{0}/\hbar $ between the two partial waves. Experimentally the
oscillation can be observed as a function of the bias voltage, which changes
the momentum of the incoming electrons. In a magnetic field the electron
trajectory becomes curved (see trajectory 2 in Fig.~\ref{Fig-model}) and the
phase difference of two partial waves mentioned above is modified as,
\begin{equation}
\Delta \phi =2p_{\mathrm{F}}r_{0}/\hbar -2\pi \Phi /\Phi _{0},  \label{dphi1}
\end{equation}%
where $\Phi $ is the magnetic flux through the projection
$S_{\mathrm{pr}}$ (see Fig.~\ref{Fig-model}) of the closed
electron trajectory onto a plane perpendicular to the vector
$\mathbf{H}$. For this reason the conductance undergoes
oscillations with a period $\Phi /\Phi _{0}$. The sign in front of
the second term in Eq.~(\ref{dphi1}) is defined by the negative
sign of the electron charge. The resulting oscillations in the
conductance $\Delta G_{1}$ (\ref{dG1}) have a nature similar to
the Aharonov-Bohm effect and are related to the quantization of
the magnetic flux through the area enclosed by the electron
trajectory. In Fig.~\ref{Fig-G_osc} the full expression for the
oscillatory part $\Delta G(H)$ of the conductance (the second term
in Eq.~(\ref{G})) is compared with the semiclassical approximation
$\Delta G_{1}(H,\rho _{0},z_{0})$, Eq~(\ref{dG1}). The long period
oscillation is a manifestation of the flux quantization effect and
is well reproduced by the semiclassical approximation. The
short-period oscillations originate from the effect of the
electron being focused by magnetic field.

In the absence of a magnetic field only those electrons that are
scattered off the defect in the direction directly opposite to the
incoming electrons can come back to the point-contact. When $H\neq
0$ the electrons move along a spiral trajectory (trajectory 1 in
Fig.~\ref{Fig-model}) and may come back to the contact after
scattering under a finite angle to the initial direction. For
example, if the defect is placed on the contact axis an electron
moving from the contact with a momentum $p_{z}=p_{\mathrm{F}}$
along the magnetic field returns to the contact when the
$z$-component of the momentum $p_{zk}=z_{0}m^{\ast }\Omega /2\pi
k$, for integer $k$. For these orbits the time of the motion over
a distance $z_{0}$ in the $z$ direction is a multiple of the
cyclotron period $T_{H}=2\pi /\Omega $. Thus,
after $k$ revolutions the electron returns to the contact axis at the point $%
z=0.$ The phase which the electron acquires along the spiral trajectory is
composed of two parts, $\Delta \phi =\Delta \phi _{1}+\Delta \phi _{2}.$ The
first, $\Delta \phi _{1}=p_{zk}z_{0}/\hbar $ is the `geometric' phase
accumulated by an electron with momentum $p_{zk}$ over the distance $z_{0}$.
The second, $\Delta \phi _{2}=\pi k(eHr_{k}^{2}/c\hbar )$ is the phase
acquired during $k$ rotations in the field $H,$ where $r_{k}=c\sqrt{p_{%
\mathrm{F}}^{2}-p_{zk}^{2}}/eH$ is the radius of the spiral trajectory.
Substituting $p_{zk}$ and $r_{k}$ in the equation for $\Delta \phi $ we find%
\begin{equation}
\Delta \phi =\pi ka_{H}^{2}/\lambdabar _{\mathrm{F}}^{2}+z_{0}^{2}/4\pi
ka_{H}^{2}.
\end{equation}%
This is just the phase shift that defines the period of oscillation of the
first term in the contribution $\Delta G_{2}$ (\ref{dG2}) to the
conductance. It describes a trajectory which is straight for the part from
the contact to the defect and spirals back to the contact by $k$ windings.
The second term in Eq.~(\ref{dG2}) corresponds to a trajectory consisting of
helices in the forward and reverse paths, with $k$ and $k^{\prime }$ coils,
respectively.

Note that, although the amplitude of the oscillation $\Delta G_{2}$ (\ref%
{dG2}) is smaller by a factor $\hbar \Omega /\varepsilon _{\mathrm{F}}$ than
the amplitude of the contribution $\Delta G_{1}$ (\ref{dG1}), the first
depends on the depth of the defect as $z_{0}^{-3/2}$ and $z_{0}^{-1}$ while $%
\Delta G_{1}\sim z_{0}^{-2}.$ The slower decreasing of the amplitude for $%
\Delta G_{2}$ is explained by the effect of focusing of the electrons in the
magnetic field.

In a high magnetic field the selection of semiclassical
trajectories that connect the contact and the defect is restricted
by the quantization condition. The projection of the momentum
$p_{z,n}$ (\ref{pz}) in the direction of the vector $\mathbf{H}$
is quantized and for a fixed quantum number $n$ $p_{z,n}$ depends
on $H$. For increasing magnetic field the distance between the
Landau levels, $\hbar \Omega $, increases and $p_{z,n}$ decreases
until $\varepsilon _{n0}=\varepsilon _{\mathrm{F}}$. As a result,
for sufficiently large $z_{0}$ each term in the conductance
(\ref{G}) corresponding to the set of quantum numbers $\left(
n,n^{\prime }\right) $ undergoes one more oscillation. This is
confirmed by the results presented
in Fig.~\ref{Fig-G_osc 0}, in which the dependencies of the $\Delta G(H)$ (%
\ref{G}) and the semiclassical asymptotic $\Delta G_{2}(H,\rho _{0}=0,z_{0})$
(\ref{dG2}) are shown for a position of the defect on the contact axis $%
(\rho _{0}=0).$

In order to observe experimentally the predicted effects it is
necessary to satisfy a few conditions: a) The distance between
Landau levels $\hbar \Omega $ is larger then the temperature
$k_{\mathrm{B}}T.$ This is the condition for observing effects of
the quantization of the energy spectrum. b) The radius of electron
trajectory, $r_{H}$, and the distance between the contact and the
defect, $r_{0}$, are much smaller then the mean free path of the
electrons for electron-phonon scattering. This condition is
necessary for the realization of the almost ballistic electron
kinetics (the scattering is caused only by a single defect) that
has been considered. c) For the observation of the
Aharonov-Bohm-type oscillations the position $\rho _{0}$
of the defect in the plane parallel to the interface must be smaller then $%
r_{H}$, i.e. the defect must be situated inside the `tube' of electron
trajectories passing through the contact. At the same time the inequality $%
\rho _{0}>a_{H}=\sqrt{r_{H}\lambdabar _{\mathrm{F}}}$ must hold in
order that a magnetic flux quantum $\Phi _{0}$ is enclosed by the
area of the closed trajectory. d) The distance $r_{0}$ should not
be very large on the scale of the Fermi wave length, because in
such case the amplitude of the quantum oscillations resulting from
the electron scattering by the defect becomes small. Although
these conditions restrict the possibilities for observing the
oscillations severely, all conditions can be realized, e.g., in
single crystals of semimetals (such as Bi, Sb and their ordered
alloys) where the electron mean free path can be up to millimeters
and the Fermi wave length $\lambda _{\mathrm{F}}\sim 10^{-8}$m.
Also, as possible candidates for the observation of predicted
oscillations one may consider the metals of the first
group, the Fermi surface of which has small pockets with effective mass $%
m^{\ast }\simeq 10^{-2}\div 10^{-3}m_{0}$ \cite{LAK}. For
estimating the periods and amplitudes of the oscillations we shall
use the characteristic values of the Fermi momentum
$p_{\mathrm{F}}$ and effective (cyclotron) mass $m^{\ast }$ for
the central cross-section of the electron ellipsoids of the Bi
Fermi surface, $p_{\mathrm{F}}\simeq 0.6\cdot 10^{-26}$kg\,m/s and
$m^{\ast }/m_{0}\simeq 0.008$ \cite{Fal}. For such parameters the
magnetic field of $
H=0.03$ in units $\lambdabar _{\mathrm{F}}/r_{H}$ shown in Figs. (\ref%
{Fig-G_osc 0}), (\ref{Fig-G_osc}) corresponds to $H\simeq 0.1$T.

The amplitude of the conductance oscillations depends mainly on
the constant of electron-defect interaction $g$ (\ref{g}), which
can be estimated using an effective electron scattering cross
section $\sim $ $\lambda _{\mathrm{F} }^{2}.$ In the plots of
Figs.~\ref{Fig-G_osc 0} and \ref{Fig-G_osc} we used a typical
value for the dimensionless constant $\widetilde{g}\sim 0.5$. The
long-period oscillations (see Fig.~\ref{Fig-G_osc}) require a
large $\rho _{0},$ the distance between the contact and the defect
in the plane of interface, and their relative amplitude is of the
order of $10^{-3}G_{c0}.$ The amplitude of short-period
oscillations for such arrangement of the contact and the defect is
small, $\sim 10^{-4}G_{c0}$, but it increases substantially and
becomes $10^{-3}G_{c0}$ if the defect is situated at the contact
axis (see Fig.~\ref{Fig-G_osc 0}). The amplitude of the
oscillations (\ref{dG0}) having de Haas-van Alphen period is
proportional
to the small parameter $\left( \hbar \Omega /\varepsilon _{\mathrm{F}%
}\right) ^{3/2},$ which for $H\sim 0.1T$ is of the order of $\sim
10^{-3}G_{c0}.$ Comparing this to previous STS experiments
\cite{Stipe}, where signal-to-noise ratios of $5\cdot 10^{-4}$ (at
1~nA, 400~Hz sample frequency) have been achieved, it should be
possible to observe the predicted conductance oscillations.

The predicted oscillations, Eqs.~(\ref{dG1}) and (\ref{dG2}), are
not periodic in $H$ nor in $1/H$. Their typical periods can be
estimated as a difference $\Delta H$ between two nearest-neighbor
maxima. For the short-period oscillations (\ref{dG2}) we find
\begin{equation}
\left( \frac{\Delta H}{H}\right) _{SP}\simeq \frac{2\lambdabar _{\mathrm{F}%
}^{2}}{a_{H}^{2}}\left( 1-\left( \frac{z_{0}\lambdabar _{\mathrm{F}}}{2\pi
a_{H}^{2}}\right) ^{2}\right) ^{-1}  \label{sp}
\end{equation}%
The period (\ref{sp}) depends on the position of the defect. It is
larger than the period of de Haas-van Alphen oscillation, $\left(
\Delta H/H\right) _{dHvA}\simeq 2\lambdabar
_{\mathrm{F}}^{2}/a_{H}^{2}.$ Both of these periods are of the
same order of magnitude as can be seen from Fig.~\ref{Fig-G_osc
0}. For a semimetal $\left( \Delta H\right) _{SP}\sim 10^{-2}$T in
a field of $H\sim 0.1$T. The characteristic interval of the
magnetic fields for the long-period oscillations is $\left( \Delta
H/H\right) _{LP}\sim 0.1$T as can be seen from
Fig.~\ref{Fig-G_osc}.

The experimental study of the magneto-quantum oscillations of the
conductance of a tunnel point-contact considered in this paper may be used
for a determination of the position of defects below a metal surface,
similar to the current-voltage characteristics considered in Ref.~\cite%
{Avotina1}. Although the dependence $G(H)$ with magnetic field is
more complicated then the dependence $G(V)$ on the applied bias,
in some cases such investigations may have advantages in
comparison with the methos proposed in \cite{Avotina1} because
with increasing voltage the inelastic mean free path of the
electrons decreases, which restricts the use of voltage dependent
oscillations.

One of the authors (Ye. S. A) is supported by the INTAS grant for Young
Scientists (No 04-83-3750) and partly supported by grant of President of
Ukraine (No. GP/P11/13) and one of the authors (Yu.A.K.) was supported by
the NWO visitor's grant.

\section{Appendix: Summation over quantum numbers in semiclassical
approximation.}

Here we illustrate the procedure for the calculations of the
correction to the conductance due to the presence of the defect in
the semiclassical approximation. At $n_{\max
}(\varepsilon_{\mathrm{F}}) \gg 1$ in the Eq.~(\ref{G}) the
summation over discrete quantum numbers $n^{\prime } $ and
$n^{\prime \prime }$ \ can be carried out using the Poisson
summation formula. Let us consider the sum of the functions
$\chi(\varepsilon _{\mathrm{F}},n,\mathbf{r}_{0}) $ (\ref{khi})
\begin{gather}
S=\sum\limits_{n=0}^{n_{\max }(\varepsilon _{\mathrm{F}})
}\chi(\varepsilon _{\mathrm{F}},n,\mathbf{r}_{0}) =S_{1}+S_{2}=
\tag{A1}  \label{S1S2} \\
\int\limits_{0}^{n_{\max }}dn\chi(\varepsilon
_{\mathrm{F}},n,\mathbf{r}_{0}) +\sum\limits_{k=-\infty ,k\neq
0}^{\infty }\left( -1\right) ^{k}\int\limits_{0}^{n_{\max
}}dn\chi(\varepsilon _{\mathrm{F}},n, \mathbf{r}_{0}) \, e^{2i\pi
kn}.  \notag
\end{gather}%
By using the Tricomi asymptotic for the Laguerre polynomials at $n\gg 1$
\cite{Tricomi} we find an expression for the first term $S_{1}$ in Eq.~(\ref%
{S1S2}) for fields that are not too high such that $n$ is large and $\rho
_{0}/(2a_{H}\sqrt{\left( 2n+1\right) )}<1$,
\begin{gather}
S_{1}\simeq \sqrt{\frac{2}{\pi }}\int\limits_{0}^{n_{\max }}\frac{dn}{\sqrt{%
\left( 2n+1\right) \sin 2\theta }}\cos \left[ \left( 2n+1\right) 2\theta
-\left( n+\frac{1}{2}\right) \left( 2\theta -\sin 2\theta \right) \right]
\times  \tag{A2}  \label{S1} \\
\exp \left( \frac{i}{\hbar }z_{0}\sqrt{2m^{\ast }\left( \varepsilon _{%
\mathrm{F}}-\hbar \Omega \left( n+\frac{1}{2}\right) \right) }\right) ,
\notag
\end{gather}%
where%
\begin{equation}
\sin ^{2}\theta =\frac{\rho _{0}^{2}}{4a_{H}^{2}\left( 2n+1\right) }.
\tag{A3}  \label{si}
\end{equation}%
For large $n$ the functions in the integrand of Eq.~(\ref{S1}) rapidly
oscillate and $S_{1}$ can be calculated by the method of stationary phase
points. As can be seen from Eq.~(\ref{si}), for $n\sim n_{\max }\sim
\varepsilon _{\mathrm{F}}/\hbar \Omega $ we have $\sin \theta \approx \rho
_{0}/2r_{H}$, where $r_{H}=v_{\mathrm{F}}/\Omega $ is the radius of electron
trajectory. For $\rho _{0}\ll r_{H}$ in Eq.~(\ref{S1}) we can make the
approximations $n\theta \sim \rho _{0}/\lambda _{\mathrm{F}},$ $n\left(
2\theta -\sin 2\theta \right) \sim \left( \rho _{0}/r_{H}\right) ^{2}\left(
\rho _{0}/\lambda _{\mathrm{F}}\right) $, and $(z_{0}/\hbar )\sqrt{2m^{\ast
}\left( \varepsilon _{\mathrm{F}}-\hbar \Omega \left( n+\frac{1}{2}\right)
\right) }\sim z_{0}/\lambda _{\mathrm{F}}.$ If $\rho _{0}$ or $z_{0}$ is
much larger than $\lambda _{\mathrm{F}},$ and the second term under the
cosine in Eq.~(\ref{S1}) is of order unity so that it can be considered as a
slowly varying function, the stationary phase point of the integral (\ref{S1}%
) is given by,
\begin{equation}
n_{\mathrm{st}}\simeq \frac{\varepsilon _{\mathrm{F}}}{\hbar \Omega }\frac{%
\rho _{0}^{2}}{r_{0}^{2}},  \tag{A4}
\end{equation}%
where $r_{0}=\sqrt{\rho _{0}^{2}+z_{0}^{2}}$ is the distance between the
point contact and the defect. The asymptotic value of $S_{1}$ takes the form%
\begin{equation}
S_{1}\simeq -\frac{ir_{H}z_{0}}{r_{0}^{2}}\exp \left( \frac{i}{\hbar }p_{%
\mathrm{F}}r_{0}-i\pi \frac{\Phi }{\Phi _{0}}\right) ,  \tag{A5}
\label{S_asym}
\end{equation}%
where $\Phi $ is given by Eq.(\ref{FI}).

The second term $S_{2}$ in the sum (\ref{S1S2}) describes an oscillation of
a different type. We consider this term for a defect position with $\rho
_{0}=0.$ Replacing the integration over $n$ by the integration over momentum
along the magnetic field $p_{n}=\sqrt{2m^{\ast }\left( \varepsilon _{\mathrm{%
F}}-\hbar \Omega \left( n+\frac{1}{2}\right) \right) }$ we rewrite the
second term in Eq.~(\ref{S1S2}) in the form
\begin{equation}
S_{2}\simeq \sum\limits_{k=-\infty ,k\neq 0}^{\infty }\left( -1\right)
^{k}\int\limits_{0}^{\sqrt{2m^{\ast }\varepsilon _{\mathrm{F}}}}\frac{%
p_{n}dp_{n}}{m^{\ast }\hbar \Omega }\exp \left[ 2\pi ki\left( \frac{%
\varepsilon _{\mathrm{F}}}{\hbar \Omega }-\frac{p_{n}^{2}}{2m^{\ast }\hbar
\Omega }\right) +\frac{i}{\hbar }p_{n}z_{0}\right] .  \tag{A6}  \label{S2}
\end{equation}%
The stationary phase points $p_{n}=p_{\mathrm{st}}$ of the integrals (\ref%
{S2}) are%
\begin{equation}
p_{\mathrm{st}}=\frac{z_{0}m^{\ast }\Omega }{2\pi k}.  \tag{A7}  \label{p_st}
\end{equation}%
Note that the stationary phase point (\ref{p_st}) exists if $k>0$ and $%
z_{0}\leq 2\pi k\,r_{H}.$ The momenta (\ref{p_st}) have a clear physical
meaning: The time $t=z_{0}m^{\ast }/$ $p_{\mathrm{st}}$ of the classical
motion of electron from the contact to the defect is an integer multiple of
the period $T_{H}=2\pi /\Omega $ of the motion in the field $H,$ $t=kT_{H}.$
This is the same condition as is applicable for longitudinal electron
focusing \cite{Sharvin}, in which case the electrons move across a thin film
from a contact on one side to a contact on the opposite surface and the
magnetic field is directed along the line connecting the contacts. The
asymptotic expression for $S_{2}$ (\ref{S2}) is given by,
\begin{equation}
S_{2}\simeq \frac{z_{0}}{2\pi a_{H}}\sum\limits_{k=\left[ z_{0}/2\pi r_{H}%
\right] }^{\infty }\left( -1\right) ^{k}\frac{1}{k^{3/2}}\exp \left( \pi ki%
\frac{a_{H}^{2}}{\lambdabar _{\mathrm{F}}^{2}}+\frac{iz_{0}^{2}}{4\pi
ka_{H}^{2}}\right) ,  \tag{A8}  \label{S2_asym}
\end{equation}%
where $\left[ x\right] $ is the integer part of the number $x.$

\


\begin{thebibliography}{99}
\bibitem{Ludoph1} B. Ludoph, M.H. Devoret, D. Esteve, C. Urbina and J.M. van
Ruitenbeek, Phys. Rev. Lett., \textbf{82,} 1530, (1999).

\bibitem{Untiedt} C. Untiedt, G. Rubio Bollinger, S. Vieira, and N. Agra{%
\"{\i}}t, Phys. Rev. B, \textbf{62}, 9962 (2000).

\bibitem{Ludoph} B. Ludoph and J. M. van Ruitenbeek, Phys. Rev. B, \textbf{61%
}, 2273 (2000).

\bibitem{Kempen} A. Halbritter, Sz. Csonka, G. Mih{\'{a}}ly, O. I.
Shklyarevskii, S. Speller, and H. van Kempen, Phys. Rev. B, \textbf{69},
121411 (2004).

\bibitem{Avotina1} Ye. S. Avotina, Yu. A. Kolesnichenko, A.N. Omelyanchouk,
A.F. Otte, and J.M. van Ruitenbeek, Phys. Rev. B \textbf{71},
115430 (2005).

\bibitem{Namir} A. Namiranian, Yu. A. Kolesnichenko, and A. N. Omelyanchouk,
Phys. Rev. B, \textbf{61}, 16796 (2000).

\bibitem{Avotina2} Ye. S. Avotina, and Yu. A. Kolesnichenko, Fiz. Nizk.
Temp., \textbf{30, } 209 (2004) [Low Temp. Phys., \textbf{30}, 153 (2004)].

\bibitem{Avotina3} Ye. S. Avotina, A. Namiranian, and Yu. A. Kolesnichenko,
Phys. Rev. B, \textbf{70}, 075908 (2004).

\bibitem{Avotina4} Ye. S. Avotina, Yu. A. Kolesnichenko, A.F. Otte, and J.M.
van Ruitenbeek, Phys. Rev. B, \textbf{74}, 085411 (2006).

\bibitem{Wend} N. Quaas, PhD thesis, G\"{o}ttingen University (2003); N.
Quaas, M. Wenderoth, A. Weismann, R.G. Ulbrich and K. Sch\"{o}nhammer, Phys.
Rev. B \textbf{69}, 201103(R) (2004).

\bibitem{LAK} I.M. Lifshits, M.Ya. Azbel', and M.I. Kaganov, \ '\textit{%
Electron theory of metals}', New York, Colsultants Bureau (1973).

\bibitem{Abrikosov} A. A. Abrikosov, \textit{'Fundamentals of the theory of
metals'}, North Holland, 1988.

\bibitem{Bogachek} E. N. Bogachek, I. O. Kulik and R. I. Shekhter, Zh. Exp.
Teor. Fiz.,\textbf{92}, 730 (1987). [Sov. Phys., JETP, \textbf{65}, 411
(1987)].

\bibitem{Bogachek1} E. N. Bogachek, I. O. Kulik and R. I. Shekhter, Solid
State Commun., \textbf{56}, 999 (1985).

\bibitem{Bogachek2} E. N. Bogachek, and R. I. Shekhter, Fiz. Nizk. Temp.,
\textbf{14}, 810 (1988) [Sov. J. Low Temp. Phys., \textbf{14}, 445 (1988)].

\bibitem{Andr} N. N. Gribov, O. I. Shklyarevskii, E. I. Ass, and V. V.
Andrievskii, Fiz. Nizk. Temp., \textbf{13}, 642 (1987) [Sov. J. Low Temp.
Phys., \textbf{13}, 363 (1987)].

\bibitem{Kittel} C. Kittel, \textit{Quantun Theory of Solids, }John Wiley$%
\And $Sons Inc., New York-London (1963).

\bibitem{KMO} I. O. Kulik, Yu. N. Mitsai, and A. N. Omelyanchouk, Zh. Exp.
Teor. Fiz., \textbf{63}, 1051 (1974).

\bibitem{LandauQM} L. D. Landau and E. M. Lifshits, \textit{Quantum Mechanics%
}, Pergamon, Oxford (1977).

\bibitem{ISh} I. F. Itskovich and R. I. Shekhter, Fiz. Nizk. Temp., \textbf{%
11}, 373 (1985) [Sov. J. Low Temp. Phys., \textbf{11}, 202 (1985)].

\bibitem{Sharvin} Yu.V. Sharvin, Zh. Exp. Teor. Fiz., \textbf{48}, 984
(1965).

\bibitem{Lifshits} I.M. Lifshits, Zh. Exp. Teor. Fiz., \textbf{32}, 1509
(1957).

\bibitem{Kosevich} A.M. Kosevich, and V.V. Andreev, Zh. Exp. Teor. Fiz.,
\textbf{38}, 882 (1960).\nolinebreak

\bibitem{Tricomi} H. Bateman, A. Erdelyi, \textit{Higher Transcendental
Functions}, V.2, Mc Graw-Hill Book Company, INC (1953).

\bibitem{Fal} L. A. Fal'kovskii, Physics-Uspekhi, \textbf{11}, 1 (1968).

\bibitem{Stipe} B. C. Stipe, M. A. Rezaei, and W. Ho, Rev. Sci. Instr.
\textbf{70}, 137 (1999).
\end{thebibliography}
\end{document}